# Ambient and high-pressure electrical transport and structural investigations of magnetic Weyl semimetal PrAlGe

U. Dutta[1*], P. Král[1,2], M. Míšek[1], B. Joseph[3] and J. Kaštil[1†]

[1]Institute of Physics, Academy of Sciences of the Czech Republic, Cukrovarnicka 10/112, 162 00 Prague 6, Czech Republic
[2]Charles University, Faculty of Mathematics and Physics, Department of Condensed Matter Physics, Ke Karlovu 5, 121 16 Prague 2, Czech Republic
[3]Elettra-Sincrotrone Trieste S.C. p.A., S.S. 14, Km 163.5 in Area Science Park, Basovizza 34149, Italy

[*]dutta@fzu.cz, [†]kastil@fzu.cz

## ABSTRACT

We present ambient and high-pressure electrical transport and structural properties of recently discovered magnetic Weyl semimetal PrAlGe. Electrical resistivity at ambient pressure shows an anomaly at $T_C = 15.1$ K related to the ferromagnetic transition. Anomalous Hall effect (AHE) is observed below $T_C$. We observe a 1.4 K/GPa increase of $T_C$ with pressure, resulting in $T_C \approx 47$ K at 23.0 GPa. Strong competition between Lorentz force and spin-scattering mechanisms suppressed by magnetic field is deduced from the magnetoresistance measurements under pressure. As in the ambient pressure case, the AHE is found to be present below $T_C$ up to the highest applied pressure. We observe a clear anomaly in the pressure dependence of $T_C$, magnetoresistance and Hall effect at 12.5 GPa suggesting the occurrence of a pressure-induced electronic transition at this pressure. X-ray diffraction (XRD) experiment under pressure revealed the lattice structure to be stable up to ~19.6 GPa with the absence of any symmetry changing structural phase transition from the initial $I4_1md$ structure. Careful analysis of the pressure dependent XRD data reveal an isostructural transition near 11 GPa. Observed isostructural transition may be related to the pressure-induced electronic transition deduced from the magnetoresistance and Hall effect data.

## INTRODUCTION

Weyl semimetals (WSM) represent a class of topologically non-trivial materials with linear electron dispersion at special points in momentum space called Weyl points in the vicinity of Fermi level. Contrary to the other topological materials such as topological insulators and topological superconductors [1–7], no gap is present in electron density of states of the WSMs. Their nontrivial band topology in momentum space leads to many novel properties such as chiral magnetic effect, ultrahigh electron mobility, negative longitudinal magnetoresistance, extreme magnetoresistance (XMR) or three-dimensional quantum Hall effect [1]. During the past few years several nonmagnetic WSMs such as TaAs class of materials (type-I) and $WTe_2$ family of compounds (type-II) were experimentally discovered with their peculiar transport properties [3-7]. WSMs are promising candidate for application in advanced electronics, e.g. in valleytronics for valley filtering [8] or in construction of super-lenses for scanning tunneling microscopes [9], which underlines the pressing needs for better understanding of variety of WSM materials and finding ways of manipulating their electronic properties.

Compared with the rapid experimental development on nonmagnetic WSMs, only a few materials have been experimentally discovered as magnetic WSMs [10-14]. The magnetic WSM materials provide an ideal platform to study the interplay between massless Dirac fermions and ordered magnetic moments. Furthermore, when magnetism is incorporated into this view of topology, various novel phenomena are to be expected [15-19]. Large intrinsic anomalous Hall conductivity (AHC) can emerge in the plane normal to the direction of the magnetization due to the associated Berry curvature that remains finite when evaluated over the Brillouin zone below the Fermi energy [20]. In contrast to the extrinsic mechanisms of anomalous Hall effect (AHE) which is related to the scattering between electrons and impurities with spin–orbit coupling [21], the intrinsic (Karplus–Luttinger) mechanism originates from an anomalous velocity resulting from a phase shift in the electronic wave-packet, which is independent of impurities [22,23]. At present, several magnetic Weyl semimetallic materials are discovered including pyrochlore iridate $R_2Ir_2O_7$ ($R$ = Y, rare earth) [10,11], ferromagnetic (FM) WSMs such as shandite structures ($Co_3Sn_2S_2$) [10] and magnetic Heusler and half-Heusler compounds such as $Co_2MnGa$ [13] and GdPtBi [14], but, only a limited number of compounds are found to exhibit such a large intrinsic AHC.

Quite recently, non-centrosymmetric PrAlGe has been proposed to host Weyl fermions in proximity to the Fermi level, making it more suitable for experimentally probing its Berry curvature properties and exploring the connection between band structure and transport properties [24-27]. PrAlGe with polar tetragonal $I4_1md$ crystal structure breaks inversion symmetry (IS) and below magnetic ordering temperature also the time-reversal symmetry. Ambient pressure magnetization and electrical transport measurements in PrAlGe confirmed FM order with magnetic moments along *c* axis below 15 K and a large AHE for field (H) along *c* axis in contrast to *a*-axis [24-27]. Isostructural LaAlGe [28] has been confirmed to be a nonmagnetic type-II WSM and CeAlGe [29] is identified as antiferromagnetic (AFM) type-II WSM without AHE. However, the Weyl point structure in $R$AlGe ($R$ rare-earth) is mainly governed by IS breaking. The magnetism and position of Fermi level with respect to Weyl nodes in $R$AlGe strongly depend on the magnetic rare earth ion $R$. As a consequence, chemical substitution or application of external pressure promise a straightforward route for tuning of magnetic ground states, AHE and Weyl node types in PrAlGe. For instance, transition from intrinsic (x ≤ 0.5) to extrinsic (x > 0.5) AHEs is observed in $PrAlGe_{1-x}Si_x$ [30] while both the end members PrAlGe and PrAlSi are FM WSMs with the same number of Weyl nodes. In addition, a crossover in the nature of the magnetic ground state is observed in $Ce_{1-x}Pr_xAlGe$ [31].

Application of hydrostatic pressure changes directly the electron density of states which results in relative shifts of characteristic energy levels. For instance, pressure-induced superconductivity has been observed in both type-I and type-II nonmagnetic WSMs [32-35]. Thus, a high-pressure study of PrAlGe is very important to investigate the continuous evolution between different magnetic ground states and the exploration of the interplay between magnetism, crystal structure and band topology and thereby motivates a series of future experiments on the other magnetic topological semimetals.

In this paper we present ambient and high-pressure electrical transport and high-pressure structural investigations of PrAlGe single-crystal. Curie temperature, $T_C$, continuously increases with the application of pressure starting at 15 K and reaching a value of ~47 K at 23 GPa, more than a three-fold enhancement induced by external pressure. Pressure-induced evolution of structural parameters up to 19.6 GPa is deduced from the high-pressure synchrotron XRD experiments.

# EXPERIMENTAL METHODS

## Synthesis

Single-crystalline samples of PrAlGe were prepared using the Al self-flux growth method. The high-purity elements (6N Al, 6N Ge, 3N Pr) were weighed in the ratio (Pr:Ge:Al) 1:1:10, inserted into the alumina crucible and then sealed into the silica glass tube under the protective argon atmosphere. The mix of elements was heated up to 1100 °C followed by 10°C/hr cooling down to 700 °C where the remaining Al flux was separated by centrifuge with silica wool as a sieve. The crystals of the desk-like shape with the planar dimensions of several millimeters were obtained. Homogeneity of prepared samples were verified by EDX analysis which showed a compositional homogeneity within the typical error of 2–3 % providing a stoichiometry of $Pr_{0.98}Al_{1.04}Ge_{0.98}$. Ambient pressure XRD measurements showed the samples to be single-phase with almost no foreign phases present. The selected single crystals of appropriate dimensions were then oriented by Laue X-ray diffraction method (see SI file Fig. S1) for further measurements.

## Ambient pressure transport measurement

Measurements of electrical resistivity and Hall resistivity were carried out in the Physical Property Measurement System (PPMS, Quantum Design) with the current along the crystallographic *a*-direction and magnetic field applied in several directions in the *ac* plane. The Hall resistivity was measured in the *a*-direction perpendicular to the direction of current and magnetic field. For ensuring the best possible geometry of contacts, the ultrasonic bonding with 25µm Al wire was used. Magnetic properties were measured with MPMS instrument (Quantum Design).

## Transport measurement under pressure

Temperature and field dependence of electrical resistance was measured under high-pressure up to 23 GPa in van der Pauw configuration. Single-crystals used for such a study had irregular geometry with approximate dimensions of $60\times50\times10$ µm$^3$ (see Fig. 1). The Merryll-Bassett type diamond anvil cell (DAC) optimized for PPMS was used for the high-pressure measurements. Fine gold wires were used as electrodes (Fig. 1). Hall measurement was performed in transverse configuration. The magnetoresistance and Hall data have been symmetrized and antisymmetrized, respectively, to remove the effect of the misalignment of the electrodes. The culet size of the used diamond anvils was 500 µm and gasket was made of stainless steel. The sample and the electrodes were electrically insulated from the metal gasket by a thin layer of $Al_2O_3$ and epoxy mixture. Powdered NaCl was used as pressure transmitting medium (PTM) to maintain a quasi-hydrostatic environment and to keep the electrodes in good contact with the sample. The sample pressure was measured by ruby luminescence method at room temperature.

## XRD measurement under pressure

High-pressure synchrotron-based powder X-ray diffraction experiment was carried out at the XPRESS beamline ($\lambda = 0.4957$ Å) of the Elettra synchrotron facility, Trieste, Italy. Few pieces of single-crystals of PrAlGe were ground into fine powder and loaded in a membrane-driven symmetric DAC. Two sets of experiments were conducted using two distinct pressure transmitting media: helium gas for pressures up to ~18 GPa, methanol-ethanol-water (MEW) (16:3:1) mixture for pressures up to ~19.6 GPa giving essentially identical results. Pressure was determined by conventional ruby fluorescence technique. 2D diffraction images were recorded on a DECTRIS PILATUS3 S 6M detector and were integrated to intensity vs 2θ diffraction profiles using the Dioptas software. To obtain the high-pressure structural parameters, LeBail refinements were performed using GSAS-II software.

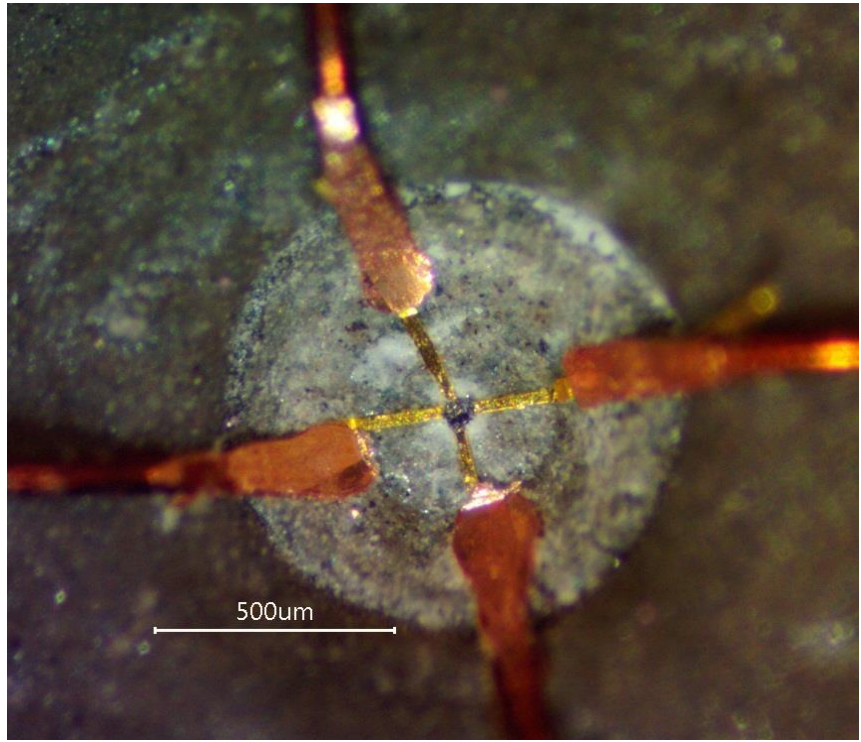

Fig. 1. Measuring configuration for resistance under high-pressure (see text for details).

## RESULTS AND DISCUSSIONS

### Ambient pressure transport measurement

Temperature and magnetic field dependence of electrical resistivity $\rho(T)$ is shown in Fig. 2. The electrical resistivity can be described by the modified Bloch-Grüneisen (BG) equation [36, 37] given below (equation 1).

$$\rho(T) = \rho_0 + A\left(\frac{T}{\theta_D}\right)^5 \int_0^{\theta_D/T} \frac{x^5}{(e^x - 1)(1 - e^{-x})} dx + c \int_{T_{0,min}}^{T_{0,max}} \exp\left(-\frac{x}{T}\right) dx \qquad (1)$$

The parameters are described below in detail. The additional exponential term $\exp(-T_0/T)$ is related to a phonon-assisted electron scattering which can be either intra-band umklapp process or inter-band scattering. Motivated by improved fit of experimental data we noticed that the final relation used in [36, 37] have some similarities to using finite integral of $\exp(-T_0/T)$ with respect to $T_0$. The parameter $T_0$ is related to the phonon-energy [38] and the integration considers the phonon-dispersion. The obtained parameter of the fit to the equation 1 are $\rho_0 = 32.31$ μΩ.cm, $\theta_D = 298.5$ K, $T_{0,max} = 70.9$ K and $T_{0,min} = 67.4$ K. The sharp change of slope of R(T) is observed at the Curie temperature ($T_C = 15.1$ K). At 2 K electrical resistivity reaches its residual value $\rho_0 = 30$ μΩ.cm leading to the residual resistivity ration RRR ≈ 3, signifying higher quality of studied samples compared to the previously published results ($\rho_0$ of the order of 100 μΩ.cm in [26]).

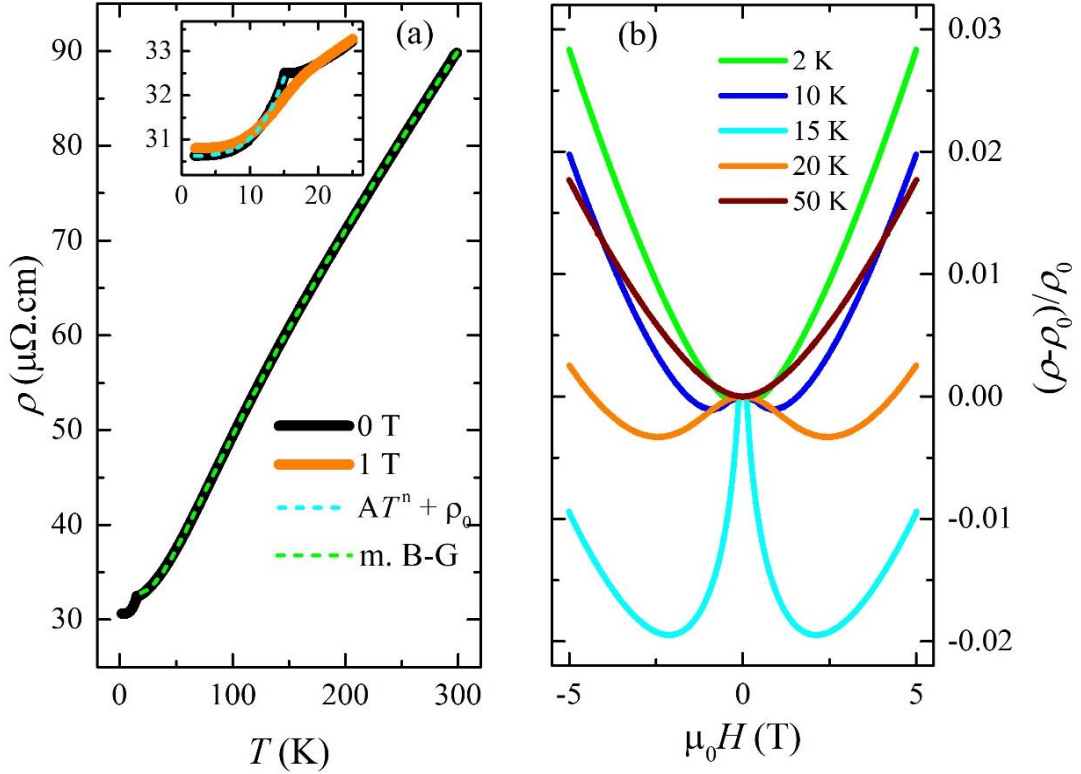

Fig. 2. (a) Temperature dependence of electrical resistivity. The inset shows the zoomed region around the Curie temperature for zero field and 1 T field applied in transversal geometry along *c*. (b) The magnetoresistance in transversal geometry of current (along *a*) and field (along *c*) at given temperatures.

Application of magnetic field of 1 T along *c*-direction results in broadening of the anomaly and shifting the transition to higher temperature (17.9 K from 15.1 K), as expected for FM materials. In the low temperature region, the resistivity data were fitted using $AT^n$ relation. Contrary to our expectations of $n = 3$ dependence [26], the best fit was obtained for $n = 3.7$. The value of the exponent of temperature dependence of resistivity of 3D Dirac semimetals was recently discussed in Ref. [39] and it was shown that the scattering time and effective density of state is temperature dependent which can result in $\rho \sim T^6$. The magnetoresistance (MR) curves measured at selected temperatures below and above $T_C$ is presented in Fig. 2 (b). The symmetric shape of MR curves with two minima presented below $T_C$ deepening and shifting to higher fields when temperature is increased towards $T_C$. For temperatures far above $T_C$ only one minimum is observed.

Results of Hall resistivity measurements at ambient pressure are reported in Fig 3. Temperature evolution of Hall resistivity was measured with external magnetic field 1 T. The significant increase, reflecting the anomalous Hall contribution, is observed below the ordering temperature and the saturated value of ≈1 μΩ.cm is reached at 2 K. The data could be used for calculation of Hall mobility using $\mu_{Hall} = R_0/\rho_{Hall}$, see inset in Fig. 3. The value $R_0 = 0.35$ μΩ.cm.T$^{-1}$ was taken as the average value of the ordinary Hall coefficients obtained for the temperatures far enough from the transition temperature, where the effects of non-saturation are negligible. The mobility is increasing with temperature up to ≈50 K and then the decreasing tendency is observed, corresponding to the growing lattice contribution to the resistivity. The field dependencies were measured at the same temperatures as the magnetoresistance data. In paramagnetic region, at T = 50 K, the measured data reflect the ordinary Hall effect with $R_0 = 0.35$ μΩ.cm.T$^{-1}$ given as the tangent of observed linear behavior. At lower temperatures, the AHE starts to influence the observed dependencies. Linear fitting of the high-field region of the data (above 0.5 T) can be used for evaluation of $R_0$ and $R_s$ values. For temperatures close to the Curie temperature, namely 15 K and 20 K, the values are affected by the non-linearity of the fitted region, as the magnetic saturation is not reached, which gives the values of $R_0$ with higher uncertainty.

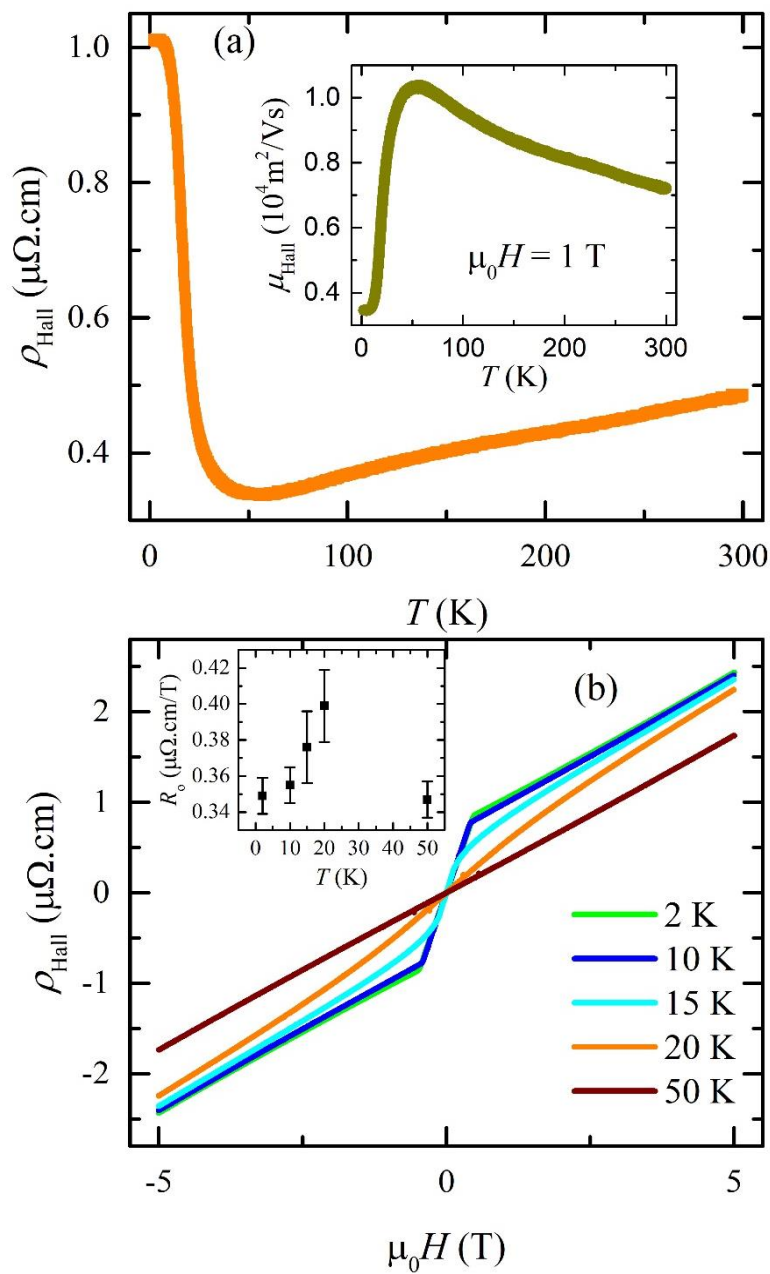

Fig. 3. (a) Temperature dependence of Hall resistivity measured in field 1 T, the inset shows calculated Hall mobility, (b) Field dependence of Hall resistivity measured at selected temperatures with the temperature evolution of Hall coefficient in the inset

**High-pressure transport measurement**

Temperature dependence of resistance ratio (R(*T*)/R(100 K)) at the respective pressures are plotted in Fig. 4 (a). The R(T) data are presented in Fig. S2 in SI file. The resistance shows metallic behavior with a low temperature drop related to the magnetic transition at all measured pressures up to 23 GPa. The resistance decreases with increasing pressure in the whole temperature range, see Fig. 4. As the pressure increases the magnetic transition becomes broadened but still clearly visible up to the highest pressure of our measurements. The critical temperature ($T_C$) is determined from the dR/d*T* plots as shown in the inset of the Fig. 4 (a). Fig. 4 (b) shows that $T_C$ continuously increases with application of pressure. The pressure of 23 GPa enhances the $T_C$ to ~47 K, which is more than three-times larger value than in ambient pressure. This huge enhancement of $T_C$ indicates that the magnetic exchange interaction between Pr moments is strongly enhanced under pressure. Such behavior was observed e.g. in $EuIn_2As_2$ [40] and $EuSn_2P_2$ [41] with the pressure-driven increase of magnetic ordering temperature attributed to the enhancement of intralayer exchange coupling by pressure. An increasing number of electrons at Fermi level mediating the Pr-Pr indirect magnetic interaction is the probable reason for the observed behavior of $T_C$ in the PrAlGe case. Pressure dependence of $T_C$ is reproducible and were observed in subsequent mounting with different piece of crystal, see green points in Fig. 4 (b) measured in close-cycle refrigerator (CCR) cryostat.

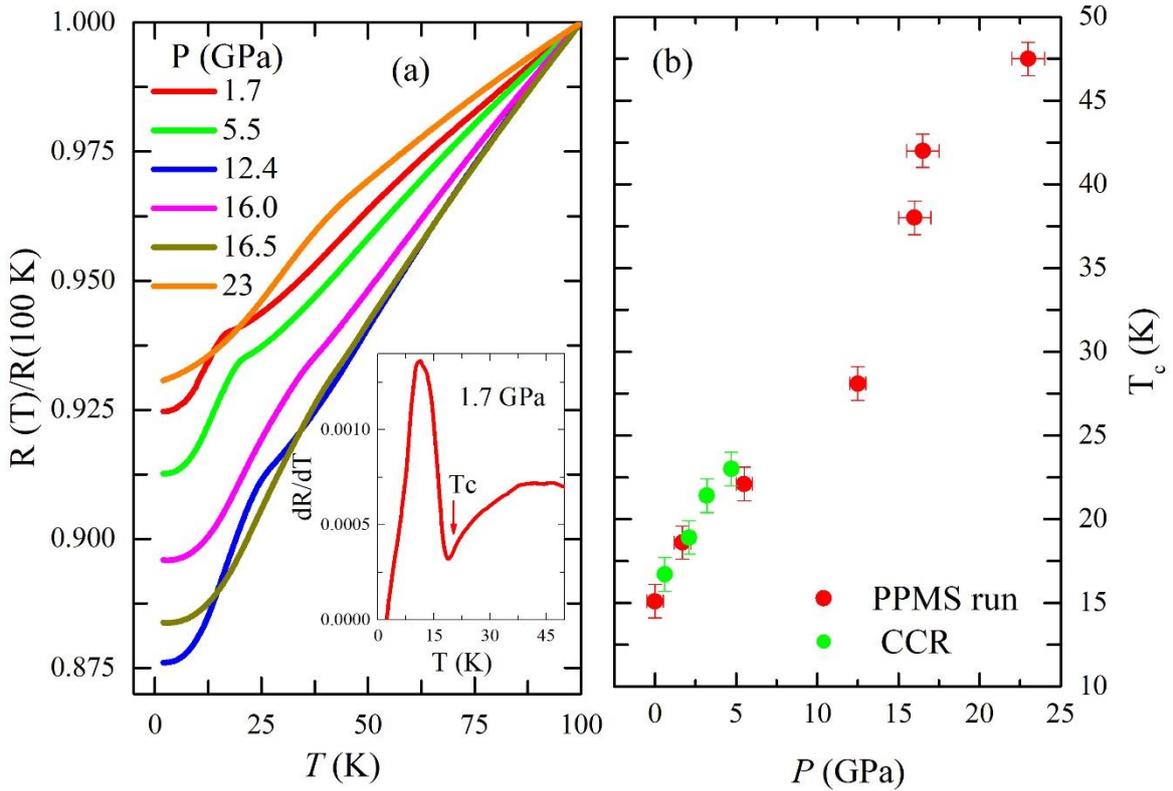

Fig. 4. (a) Temperature dependence R(*T*)/R(100 K) at various pressures up to 23 GPa. The inset shows the temperature dependence dR/d*T* at 1.7 GPa to illustrate how $T_C$ was determined at each pressure. (b) Pressure dependence of $T_C$ at two different measurements.

At low temperatures, below 60 % of $T_C$, R(T) data are fitted with the power law $R = R_0 + AT^n$ at each pressure point, see the fitting at 1.7 GPa and at 23 GPa in Fig. 5 (a) and (b). At 1.7 GPa $n \approx 2.8$ and with increasing pressure it continuously decreases as shown in the Fig. 5 (b). However, with increasing pressure the resistance of the sample develops an upturn below 10 K which cannot be correctly described by the used power-law. The ambient pressure value of $n \approx 3.7$ decrease under application of pressure and reaches value close to 2 characteristics for Fermi liquid behavior. The value of $R_0$ is continuously decreasing with increasing pressure.

Fig. 6 (a) - (d) show the electrical resistance as a function of pressure at different fields of 0 T, 0.1 T and 1 T at 1.7 GPa, 12.5 GPa, 16 GPa and 23 GPa, respectively. At 1.7 GPa, resistance exhibits very weak field dependence, however it becomes more pronounced as the pressure is increased.

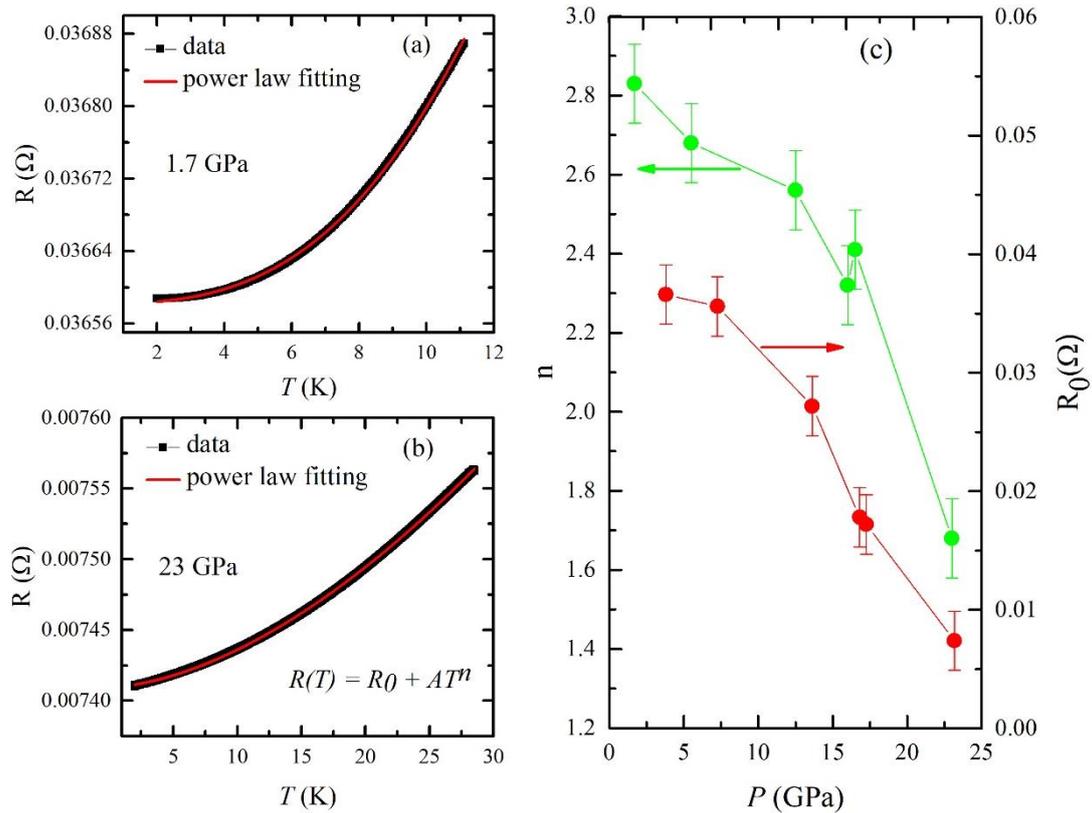

Fig. 5. (a), (b) Power law fitting of the resistance at low temperature at 1.7 GPa and 23 GPa respectively. (c) Pressure dependence of the residual resistance and the power exponent $n$ in the power-law fit of $R(T)$ below 60 % of $T_C$.

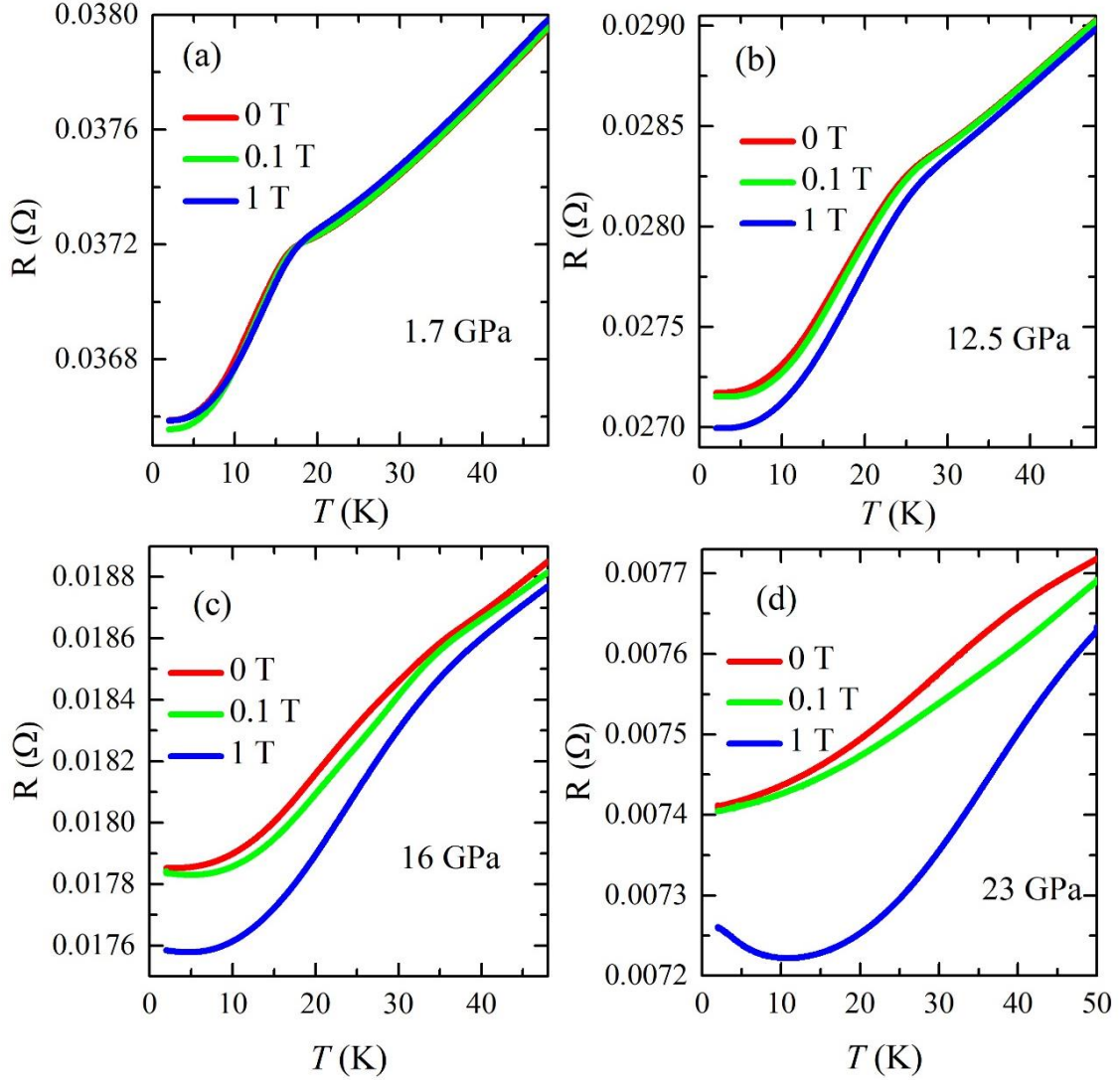

Fig. 6. (a) Temperature dependence of resistance at different fields at (a) 1.7 GPa (b) 12.5 GPa (c) 16 GPa and (d) 23 GPa

Figure 7 (a) and (b) show the field dependence of transverse magneto-resistance MR = [R(H) − R(0)]/R(0) and Hall Resistance, respectively, at 5.5 GPa at various temperatures. All the measurements were performed in the van der Pauw configuration. The MR is taken according to MR$_{sym}$(H) = [MR(μ$_0$H) + (MR(−μ$_0$H)]/2. Compared to the ambient pressure MR we observe an order of magnitude reduced MR which can be related to Fermi surface modification under application of high-pressure [42]. However, we cannot rule out effect of non-hydrostatic condition in solid pressure transmitting medium and misalignment of the sample in DAC with respect to ambient measurement. At 50 K (sufficiently far from $T_C$) the MR is positive but when $T$ is close to $T_C$, negative magnetoresistance is observed due to suppressed spin scattering by magnetic field similarly to the ambient pressure. At 2 K, the sign of MR changes back to positive because of the dominant role of Lorenz force. At high temperatures (T > $T_C$), R$_{xy}$(μ$_0$H) shows linear filed dependence as shown in Fig. 7 (b), while below $T_C$, R$_{xy}$(μ$_0$H) shows an AHE similar to the ambient pressure. Positive Hall resistance at positive magnetic field indicates the domination of hole-like carriers in PrAlGe.

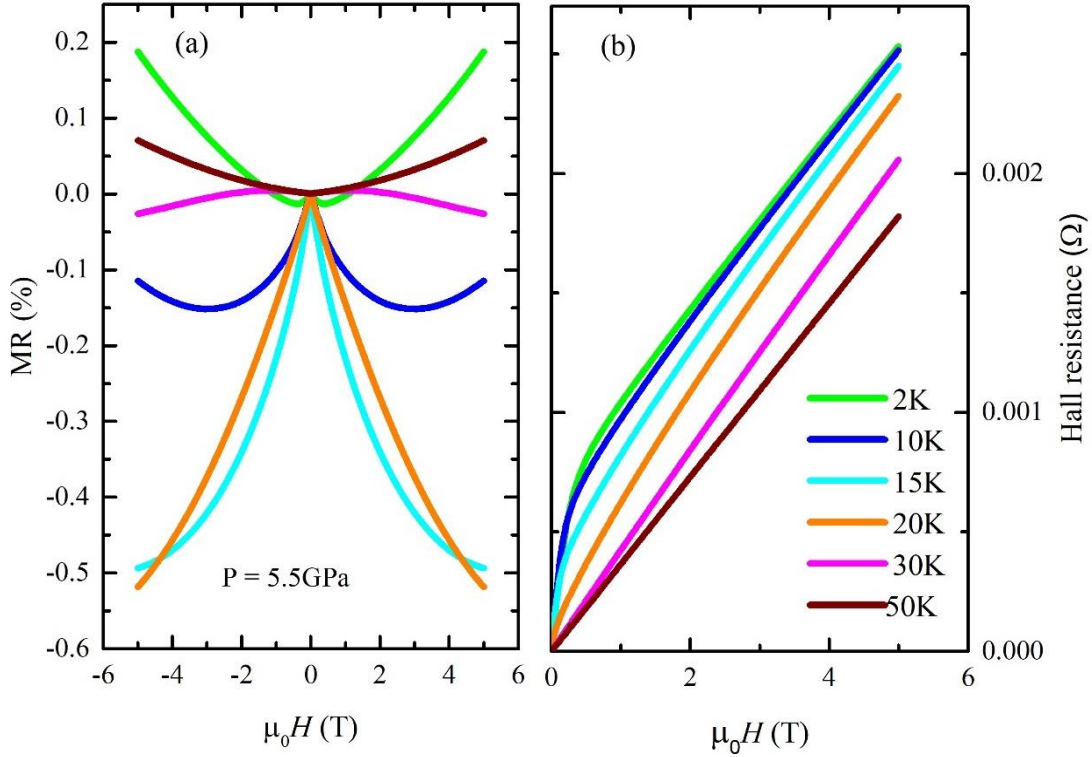

Fig. 7. (a) Field dependence of MR at 5.5 GPa at various temperatures (b) Field dependence of Hall Resistance at 5.5 GPa at at various temperatures

Figure 8 (a) and (b) show the field dependence of MR and Hall resistance, respectively, at 2 K at various pressures. At 1.7 GPa, MR shows pure quadratic behavior suggesting that the Lorenz force is controlling the MR process. At 5.5 GPa and 12.5 GPa, negative magnetoresistance is observed at low field but the data shows quadratic dependence at high field with a minimum near ∼1 T. At 16 GPa, the minimum shifts to lower field with very weak negative magnetoresistance and at 16.5 GPa MR again shows quadratic field dependence. These observations suggest a strong competition between Lorentz force mechanism and suppressed spin-scattering mechanism under pressure which will be strongly influenced by increasing magnetic interaction. At 23 GPa MR deviates from quadratic field dependence toward linear field dependence. From Fig. 8 (b), it is clear that the anomalous Hall resistance initially increases up to 16 GPa and starts to decrease above this pressure. Such behavior can suggest that Weyl-point is crossing the Fermi level at this pressure range. The clear anomaly of pressure dependence of both MR and Hall resistance at 12.5 GPa signifies electronic transition. The as-measured Hall resistance develop pronounced hysteresis, clearly observed at 23 GPa, in magnetic field which suggest that the coercive field of PrAlGe is also significantly enhanced by pressure.

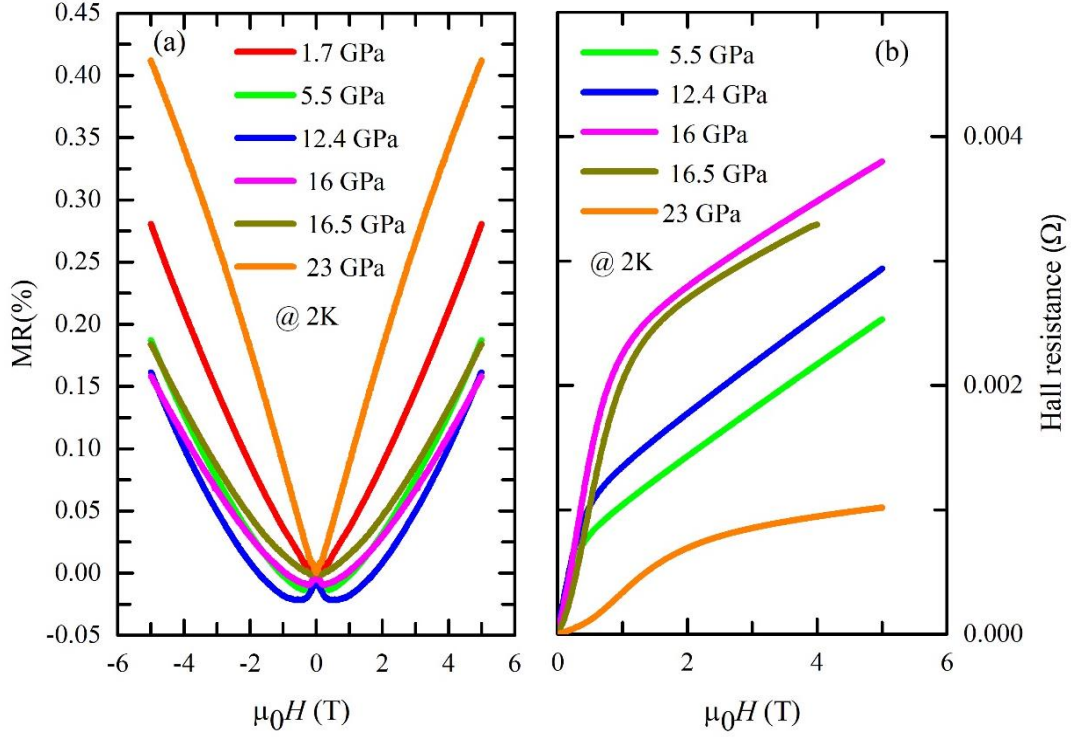

Fig. 8. (a) Field dependence of MR at 2 K at various pressures (b) Field dependence of Hall resistance at 2 K at various pressures

Ordinary Hall resistivity increases with application of pressure up to 12.5 GPa and then it starts to decrease, see Fig. 8 (b). Ordinary Hall resistivity also indicates that under application of 23 GPa, carrier density reduces to its 1/2 value. With the assumption of contribution of both electrons and hole to the conduction the result is consistent with increasing role of electrons at the Fermi level and could suggest crossover to dominant electron regime at significantly higher pressures.

Fig. 9 shows the temperature dependence of Hall resistance at various pressures at 1 T. At 5.5 GPa and 12.5 GPa Hall resistance exhibits a sharp upturn near $T_C$ followed by saturation at low temperature similarly to the ambient pressure. At 16 GPa and higher pressures, Hall resistance exhibits similar upturn near $T_C$ although followed by a maximum at lower temperatures. This observation also indicates a possible electronic transition near 12.5 GPa.

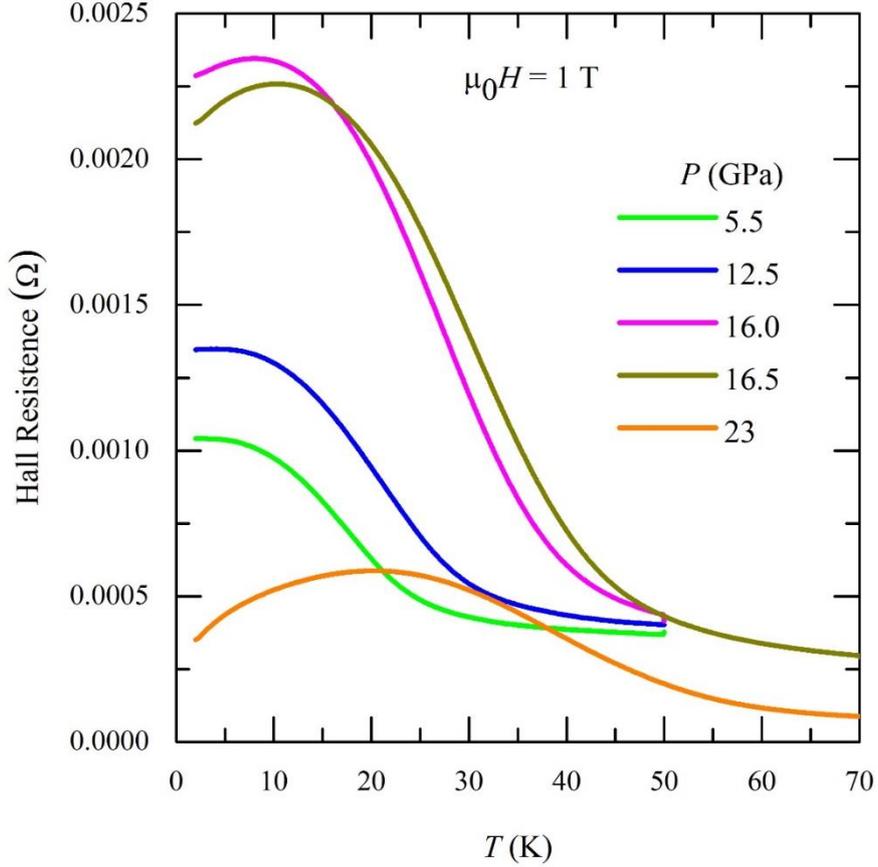

Fig. 9. Temperature dependence of Hall resistance at different pressures at 1 T.

Previous quantum-mechanical calculation reveals the existence of multiple Weyl nodes in the vicinity of Fermi level of PrAlGe [43]. Three types of Weyl nodes are present in nonmagnetic state due to the breaking of inversion symmetry. The position of the Weyl points is changed in the magnetic state and a new Weyl point W4 appears about 20 meV below the Fermi level as a result of breaking time reversal symmetry. The Weyl points closest to the Fermi level W3 and W4 are of type-I [43]. The effect of pressure on the position of Weyl nodes is rather complex and was studied e.g. for LaAlSi crystal up to 90 GPa [44]. The LaAlSi system has the same crystal structure as PrAlGe. Moreover, the previous results on related *R*AlGe and *R*AlSi compounds [45-47] suggest that the Weyl points structure is very robust and governed mainly by the crystal structure with magnetic moment of *f*-electrons localized on rare-earth atoms introduce additional Zeeman energy. The observed isostructural transition can be explained by Fermi level moving with respect to Weyl points. The decreasing ordinary Hall effect suggest possible change from dominant hole-carrier to dominant electron-carrier regime at pressures above 60 GPa. Recently, the possibility of tuning the AHE in topological semimetals has been demonstrated by chemical doping where a transition of the AHE from an intrinsic (x ≤ 0.5) to an extrinsic regime (x > 0.5) is observed in PrAlGe$_{1-x}$Si$_x$. When x is increasing in PrAlGe$_{1-x}$Si$_x$ a smaller hole and larger electron Fermi surfaces are observed as found from the sign change of the Hall resistivity. In our measurements hole-like carriers dominates over the entire pressure range as no sign change in the Hall resistivity is observed and the intrinsic Hall effect dominates over the extrinsic one up to the highest pressure similar to the ambient pressure. AHE is present up to the highest pressure of our measurements unlike the case of half-Heusler AFM GdPtBi where AHE disappears at a small pressure of 1.5 GPa [48]. The enhanced magnetic transition temperature with well localized Pr magnetic moments points to considerable increase of magnetic exchange interaction due to the increasing number of electrons mediating magnetic interaction.

**High-pressure XRD measurement**

We do not observe any sign of crystal-symmetry changing structural phase transitions up to the highest applied pressure which confirms the PrAlGe *I4$_1$md* structure is robust under pressure. A similar observation is obtained from an independent high-pressure run using methanol-ethanol-water (MEW) (16:3:1) mixture as PTM [see Fig. S4(a)]. In order to follow the pressure dependence of lattice parameters, the LeBail refinement using GSAS-II software was carried out. Figure 10 (b) shows the result of the LeBail refinement at 0.2 GPa. Due to the spotty nature of the data, there might be some preferred orientations generated by pressure which prevent us from performing the Rietveld refinement needed to comment about the bond distances and bond angles. Our LeBail analysis demonstrate that the *I4$_1$md* space group can provide a good description of the data up to the highest pressure in this study.

To obtain further input on the structural evolution of PrAlGe, one can look into the unit-cell volume vs pressure plot, which is shown in figure 10 (c). The pressure evolution of the lattice parameters obtained from the two runs are shown in Fig. S5. The unit-cell volume vs pressure plot shows a smooth evolution, however, we noticed that the entire data-set cannot be described by a single equation of state (EOS). In Fig. 10 (c) we have also presented a third-order Birch-Murnaghan equation of state (EOS) fit to the experimental data (solid red and green lines). The third-order Birch-Murnaghan equation (equation 2) is given below:

$$P(V) = \frac{3B}{2}\left[\left(\frac{V}{V_0}\right)^{\frac{7}{3}} - \left(\frac{V}{V_0}\right)^{\frac{5}{3}}\right]\left[1 + \frac{3}{4}(B' - 4)\left(\left(\frac{V}{V_0}\right)^{\frac{2}{3}} - 1\right)\right] \qquad (2)$$

where $V$ is the unit cell volume at pressure $P$, $V_0$, $B$ and $B'$ represent unit cell volume at ambient conditions, bulk modulus and its 1$^{st}$ derivative, respectively. As shown in Fig. 10 (c), we require two distinct EOS curves with varying $B$ to the describe the pressure evolution of the entire data-set. The data below 11 GPa provided a $B$ value of 81.4 GPa, which is increased to 94 GPa for the high-pressure phase above. Such a change was also confirmed by the pressure run using the methanol-ethanol-water (MEW) (16:3:1) mixture PTM [Fig. S4 (c)]. Incidentally, distinct slopes can be observed also for the out-of-plane to in-plane lattice parameter ratio (c/a) for above and below 11 GPa (Fig. S5). These observations indicate the presence of a pressure-induced isostructural phase transition close to 11 GPa. This transition is related to the electronic transition near 12.5 GPa as observed from the transport measurements. Thus, this transition may be classified as an isostructural electronic phase transition as our XRD data rules out any structural phase transition up to ~18 GPa. In comparison with the isostructural nonmagnetic type-I TaAs family of WSMs (TaAs, TaP, NbAs, NbP), our observation of a pressure-induced isostructural transition is very similar to the case of NbAs and NbP but very different from the other compounds [32,49,50]. Pressure-induced crystalline-to-amorphous phase transition has recently been observed in various magnetic topological materials such as EuIn$_2$As$_2$ [40] and EuSn$_2$P$_2$ [41] at 17 GPa and 36 GPa, respectively. We also remark that the EuIn$_2$As$_2$ and EuSn$_2$P$_2$ have much smaller B from the EOS fit (respectively around 32.3 GPa [40] and 58 GPa [41]) compared to PrAlGe. Pressure-induced rhombohedral to monoclinic phase transition is also identified at 14 GPa on a recently discover magnetic topological insulator EuSn$_2$As$_2$ [51]. Contrary to that, PrAlGe structure is relatively robust under pressure and higher-pressure structural investigation is needed to confirm any pressure-induced structural phase transition or amorphization in PrAlGe.

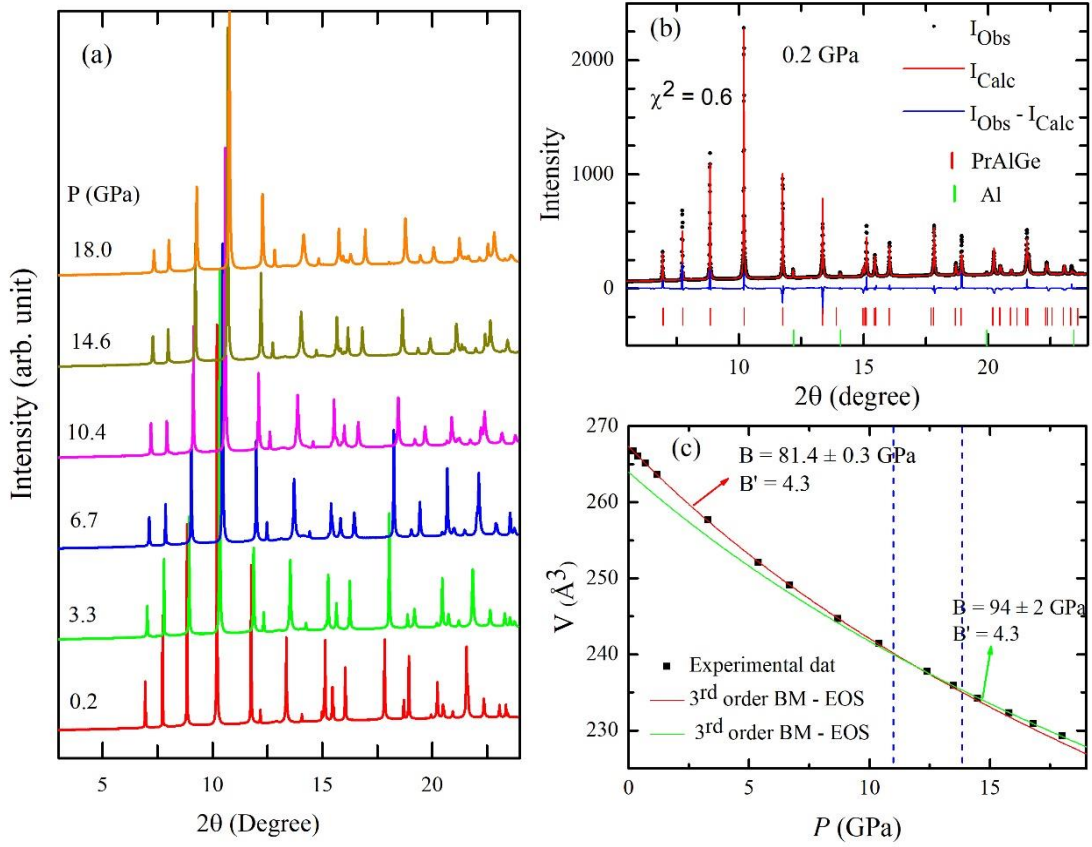

Fig. 10 (a) Powder X-ray diffraction patterns of PrAlGe at selected pressures. (b) LeBail refinement of the X-ray diffraction profile at 0.6 GPa. (c) Unit cell volume as a function of pressure. Vertical lines represent the range of pressures where the iso structural phase transition occurs.

## CONCLUSIONS

In conclusion, we have studied electrical transport properties of a recently discovered magnetic WSM, PrAlGe, at ambient pressure as well as under the application of external pressure. Crystal structure response to the applied-pressure has also been studied. More than three-fold of enhancement in critical temperature is observed under application of external pressure up to 23 GPa. This result suggests that the application of high-pressure is an effective way to enhance the magnetic transition temperature in magnetic topological materials which may help to understand the novel quantum phenomena at elevated temperatures. No pressure-induced superconductivity was observed within the experimental pressure and temperature range. Similar to the ambient pressure, AHE is present below the critical temperature with weak pressure dependence up to the highest pressure of our measurements. A pressure-induced electronic transition has been observed near 12 GPa. High-pressure XRD measurements reveal that this transition is an isostructural electronic transition, as the *I41md* structure remains stable up to ~19.6 GPa. Thus, our findings provide important insights into the interplay between electronic and structural properties in PrAlGe. These studies call for theoretical investigations to uncover the origin of the pressure-induced electronic transition in PrAlGe. Our results further reveal the potential of high-pressure to tune the magnetism and AHE on the magnetic topological materials.

# ACKNOWLEDGMENTS

Experiments were performed in MGML (mgml.eu), which is supported within the program of Czech Research Infrastructures (project no. LM2023065). High-pressure powder X-ray diffraction (PXRD) experiments have been conducted at the Elettra synchrotron facility in Trieste, Italy, as part of the research proposal with ID No. 20220618. One of us (B.J) thanks Elettra Sincrotrone for the internal project Xpress_Partner and Dr. Alain Polian for a one-week training on GLS gas-loading system.

# AUTHOR CONTRIBUTIONS

JK designed and supervised the project. PK and JK prepared samples, performed the ambient pressure transport experiments. UD, JK and MM, performed the high-pressure transport experiments. UD, PK and BJ performed the high pressure XRD experiment. PK and UD analyzed the high pressure XRD data. All the authors contributed to the manuscript writing.

# COMPETING INTERESTS

The authors declare no competing interests.

# DATA AVAILABILITY

Data supporting the findings are presented within the article and supplementary file. Access to the data file can be provided by the corresponding authors upon request